\documentclass{svjour3} 
\smartqed
\usepackage{graphicx}
\usepackage{dcolumn}% Align table columns on decimal point
\usepackage{bm}% bold math
\usepackage{color}
\usepackage{amsbsy}

\begin{document}

\title{Effects of Interplanetary Dust on the LISA drag-free Constellation}

\author{Massimo Cerdonio, Fabrizio De Marchi, Roberto De Pietri, Philippe Jetzer, Francesco Marzari, Giulio Mazzolo, Antonello Ortolan and Mauro Sereno}

\institute{M. Cerdonio and F. Marzari \at Department of Physics, University of Padova and INFN Padova, via Marzolo 8, I-35131 Padova, Italy\\    
\and
F. De Marchi \at Department of Physics, University of Trento and INFN Trento, via Sommarive 14, I-38123 Povo (Trento), Italy\\       
\and  
R. De Pietri \at Department of Physics, University of Parma and INFN Parma, I-43100 Parma, Italy\\        
\and
Ph. Jetzer \at Institute of Theoretical Physics, University of Z\"urich, Winterthurerstrasse 190, CH-8057 Z\"urich, Switzerland\\
\and
G.Mazzolo* \at Max Planck Institut f\"ur Gravitationsphysik, Callinstrasse 38, 30167 Hannover, Germany\\
*To whom address correspondence: giulio.mazzolo@aei.mpg.de 
\and
A. Ortolan \at INFN Laboratori Nazionali di Legnaro, Viale dell'Universit\`a 2, I-35020 Legnaro (Padova), Italy\\
\and
M. Sereno \at  Institute of Theoretical Physics, University of Z\"urich, Winterthurerstrasse 190, CH-8057 Z\"urich, Switzerland and Department of Physics, Politecnico di Torino, Corso Duca degli Abruzzi 24, 10129 Torino, Italy\\
}

%\address{$^1$ Department of Physics, University of Padova and INFN Padova, I-35100 Padova, Italy}

%\address{$^2$ Department of Physics, University of Trento and INFN Trento, I-38100 Povo (Trento), Italy}

%\address{$^3$ Department of Physics, University of Parma I-43100 Parma, Italy}

%\address{$^4$ Institute of Theoretical Physics, University of Z\"urich, 8057 Z\"urich, Switzerland}

%\address{$^5$ INFN Laboratori Nazionali di Legnaro, I-35020 Legnaro (Padova), Italy}

\maketitle

\begin{abstract}
The analysis of non-radiative sources of static or time-dependent gravitational fields in the Solar System  
is crucial to accurately estimate the free-fall orbits of the LISA space mission. In particular, we take into 
account the gravitational effects of Interplanetary Dust (ID) on the spacecraft trajectories. 
The perturbing gravitational field has been calculated  for some ID density distributions that fit 
the observed zodiacal light. Then we integrated the Gauss 
planetary equations to get the deviations from the LISA keplerian orbits around the Sun.
This analysis can be eventually extended to Local Dark Matter (LDM), as gravitational fields are expected to be similar for ID and LDM distributions. Under some strong assumptions on the displacement 
noise at very low frequency, the Doppler data collected during the whole LISA mission could provide 
upper limits on ID and LDM densities. 
\end{abstract}

%\pacs{04.80.Nn, 95.55.Ym}

\section{Introduction}
\label{intro}

LISA (Laser Interferometer Space Antenna) is a joint space mission by ESA and NASA which is planned to be launched 
at the end of the next decade. It consists of three identical free-falling satellites, orbiting around the Sun and 
marking the vertices of a nearly equilateral triangle with  $\simeq 5 \cdot 10^{6}$ km (1/30 AU) sides, 
located $20^{\circ}$ behind the Earth and lying in a plane that makes an angle of $60^{\circ}$ with the 
ecliptic \cite{LISA}. LISA target is the detection of gravitational waves (GWs) through the measure 
of the relative and differential motions between the spacecrafts. Among the astrophysical goals of LISA is to detect GWs  
originated by events like black holes coalescence or capture of 
compact objects by black holes \cite{Bender}. This requires a strain sensitivity of 
$10^{-20} \ \mbox{Hz}^{-1/2}< S_h^{1/2} < 10^{-17} \ \mbox{Hz}^{-1/2}$ in the $10^{-4}$ to $10^{-1}$ Hz frequency band range.

However, LISA reference masses interact also with time-dependent (e.g. planets and their satellites, 
quadrupolar pulsation of the Sun, etc.) and static (e.g. ID and LDM) local gravitational fields, 
and therefore they depart from ideal unperturbed orbits around the Sun.  
In this paper, we focused on the perturbing effects caused by the static components ID and LDM. 
As the discriminating feature between ID and LDM concerns only their coupling with the electromagnetic field, 
we expect that ID and LDM will produce similar gravitational forces on reference masses \footnote{This is correct only in Newtonian physics: in fact, DM and ID are expected to be gravitationally bound to the Galaxy and to the Solar System respectively; therefore, while gravitoelectric effects are identical for the two, that would not be the case for the gravitomagnetic ones, i.e., effects depending on the speed of the considered objects.}.
In fact, only ID reflects the solar light and can be studied through observations of zodiacal light \cite{Giese}, 
while LDM is ``dark" and its presence can be investigated by means of gravitational perturbations 
induced on orbiting bodies. It is worth mentioning that other possibilities for 
detection of DM are the searches for particular electro-weak processes beyond the Standard Model at particle 
colliders. Thus we assume that ID and LDM induce identical gravitational effects on LISA orbits, 
with the irrelevant difference that the first is luminous and the second is dark.
  
As we will show, the linear perturbation theory holds, being other gravitational forces acting on LISA reference masses much smaller than the Sun pull; under this approximation, small perturbations to ideal keplerian motions induced by different sources in the Solar System can be studied  separately.

\section{Interplanetary dust}
\label{id}

The ID cloud is a dust composed by grains with typical sizes of $10-100\  \mu$m pervading the Solar System. 
%Sources of ID particles should be  but also produce thermal emission, which is the most prominent 
The distribution of ID particles is studied by the observations of solar light scattering (i.e. the zodiacal light confined to the ecliptic plane) and by their thermal emission, which is the dominant component of night-sky light in the $5-50\ \mu$m wavelength \cite{Levasseur}. %(Levasseur-Regourd, A.C. 1996). 
Physical models of ID density distribution $\rho$ should account for its symmetries, that can be easily 
described in the Solar System Baricentric (SSB) reference frame $(x, y, z)$: i) invariance under rotation 
about the $z$ axis, and ii) invariance under reflections in the $(x, y)$ ecliptic plane \footnote{The distribution is not symmetric with respect to the ecliptic plane, but to one defined by the total angular momentum of the Solar System. The differences between the two planes are negligible for our scopes.}.
%here (x,y,z) are the coordinates of the SSB (Solar System Baricentric) reference frame. 
Usually one also assumes a static distribution and a power-law radial profile. 
As a consequence, all the proposed models factorize $\rho$ into two functions \cite{Giese}    
%In all models, the separability of $\rho(r, \beta_{\odot})$ into two factor is assumed, the first being a power $\alpha$ of solar distance $r$ and the second depending on the helioecliptic latitude $\beta_{\odot}$ \cite{Giese}, i.e., 
\begin{equation}
\rho(r, \beta_{\odot}) = \rho_0 \left(\frac{r_0}{r} \right)^{\alpha} f \left(\beta_{\odot} \right),
\end{equation}
where $r$ is the distance from the Sun, $\beta_{\odot}$ is the helioecliptic latitude, 
$\rho_0\simeq 9.6 \times 10^{-20}$ kg/$\mbox{m}^3$ is the density value 
at $r_0=1$ AU (Earth's orbit), as estimated by the integration of the meteoroid mass distribution far away from the Earth, in order to avoid its gravitational attraction on ID particles, and $f$ is a given function. 
The typical value of the parameter $\alpha$, determined from zodiacal light photometer measurements on
Helios 1 and 2 \cite{Grun}, is $1.3$.

The expression of $f \left(\beta_{\odot} \right)$ is still uncertain; however, the analytical formula which best 
reproduces the observations of zodiacal light is the so-called \textsl{ellipsoid model}    
\begin{equation} \label{eqn: model}
\rho(r, \beta_{\odot}) = \rho_0 \left(\frac{r_0}{r} \right)^{\alpha} \frac{1}{\left[1 + \left(\gamma_E \sin \beta_{\odot} \right)^2 \right]^{\alpha/2}},
\end{equation}
where $\gamma_E=\sqrt{a^2-b^2}/b$, $a$ and $b$ are the semi-major and semi-minor axes of an oblate ellipsoid 
respectively. Beyond $\sim 3$ AU in the ecliptic plane and $\sim 1.5$ AU off the ecliptic plane, no reliable density values can be obtained from the zodiacal light. Therefore we assume $a=3$ AU and $b=1.5$ AU for the semi-axes and $\gamma_E=\sqrt{3}$ \cite{Giese}. It is worth noticing that the total ID mass amount inside the considered region is of the order of $10^{17}$ kg, i.e., $\approx 10^{-8} \ M_{\oplus}$ .

The ellipsoid model of ID density depends on two parameters, $\alpha$ and $\gamma_E$, and, in order to 
numerically study gravitational effects of ID on LISA orbits, we will consider four 
cases: i) \textsl{spherical homogeneous distribution} ($\alpha=0,\gamma_E=0$), ii) \textsl{spherical distribution with power-law 
density profile}  ($\alpha=1.3,\gamma_E=0$), iii) \textsl{ellipsoidal homogeneous distribution} 
($\alpha=0,\gamma_E=\sqrt{3}$), and iv) \textsl{ellipsoidal distribution with power-law density 
profile} ($\alpha=1.3,\gamma_E=\sqrt{3})$.

\section{Method}
\label{method}

\begin{figure}[!hbt]
\begin{center}
\begin{tabular}{cc}
 \includegraphics[width = 5.9 cm, height = 5.0 cm]{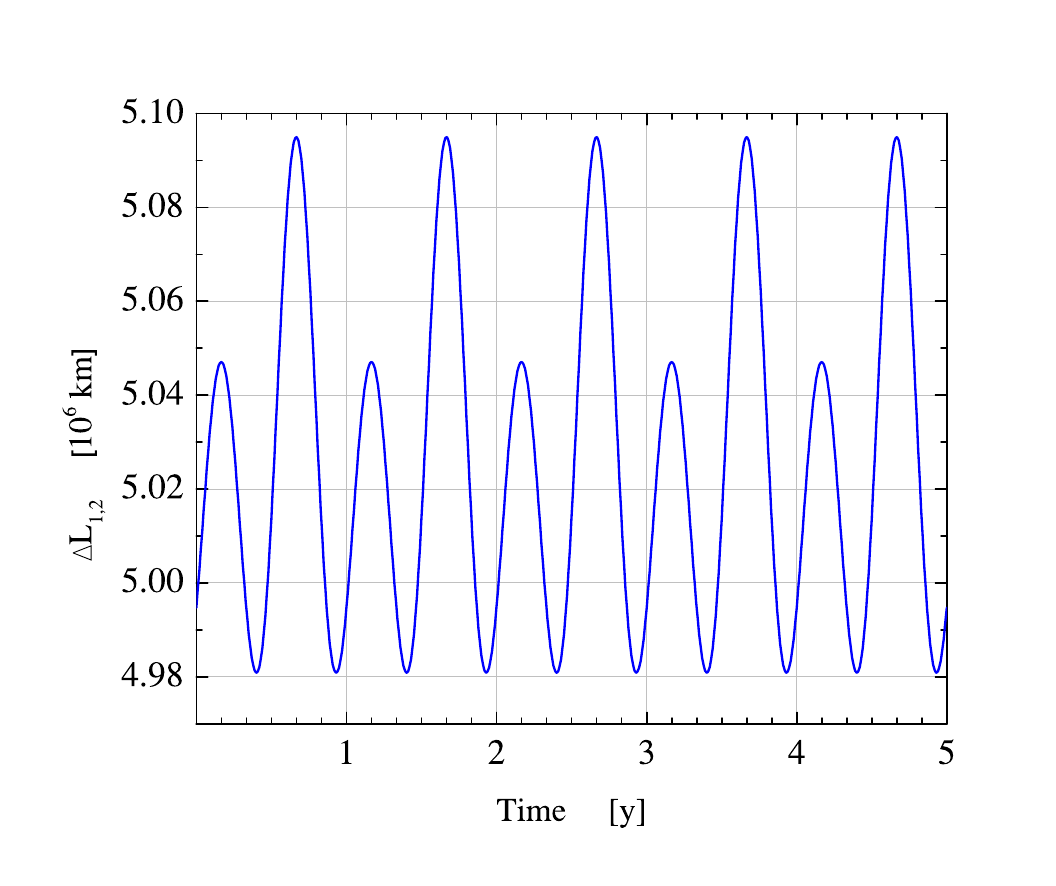}&
 \includegraphics[width = 5.9 cm, height = 5.0 cm]{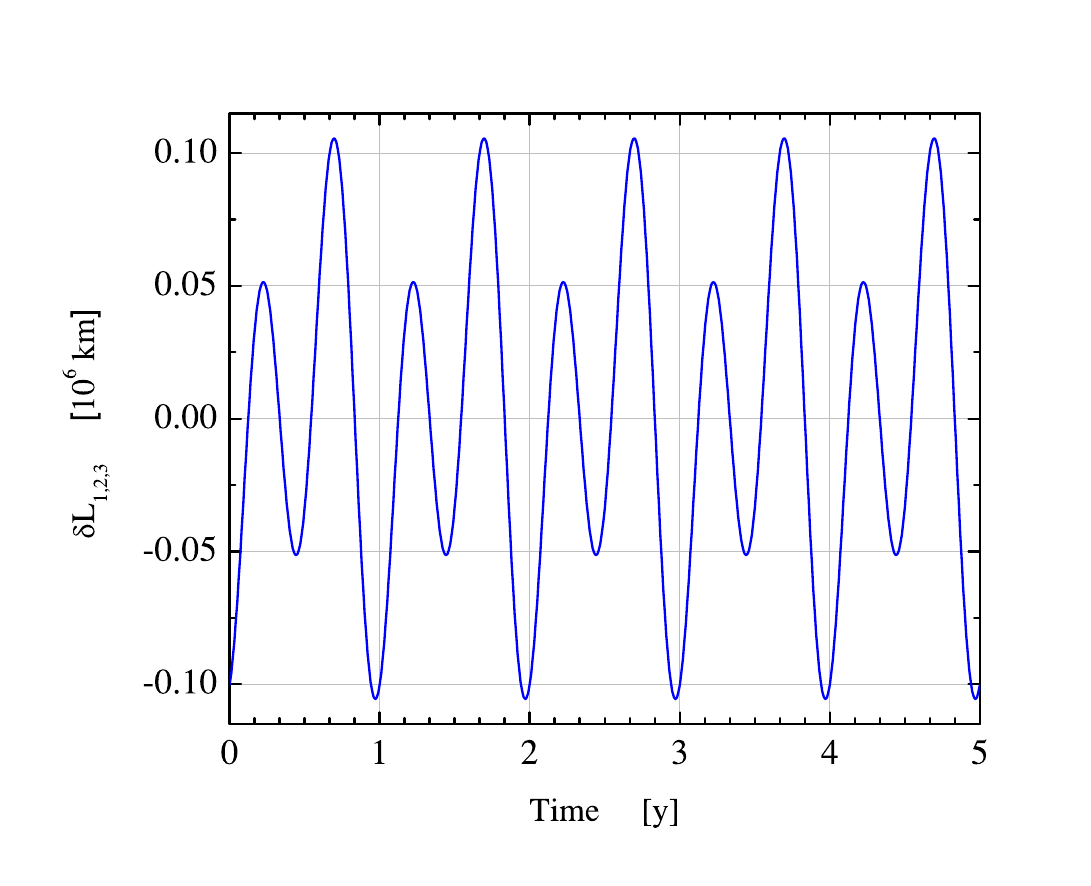}
\end{tabular}
\caption{Left, relative motion of 1 and 2 reference masses $\Delta L_{1,2}$ as a function of time. Right, differential relative motion $\delta L_{1,2,3}$ between couples 1-2 and 2-3 of reference masses as a function of time. For different couples, plots are identical but shifted in time.}
\label{variazione_distanze_imp}
\end{center}
\end{figure}

\begin{figure}[!hbt]
\begin{center}
\includegraphics[width = 8.9 cm, height = 7.2 cm]{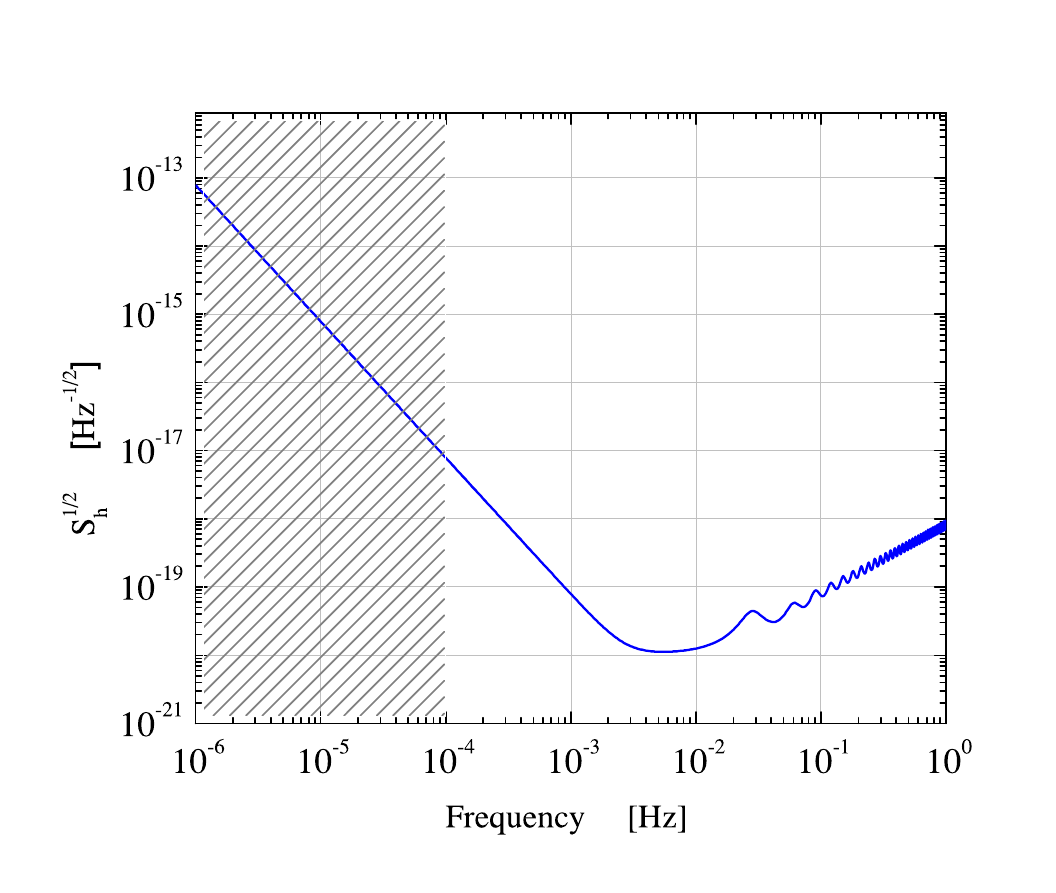}
\caption{LISA sensitivity curve as a function of frequency. The curve in the dashed region 
is calculated from ref. \cite{Bender2}, where it is discussed the sensitivity curve up to $10^{-6}$ Hz.} 
\label{sensitivity}
\end{center}
\end{figure}

As the first step we define the unperturbed LISA orbits\footnote{In this paper we neglect post-Newtonian and relativistic effects.}. The cartesian coordinates of each LISA reference mass ($k=1,2,3$) are related to the keplerian orbital elements defined in the SSB through the following equations:

\begin{equation} \label{eqn: orbita}
\left\{
\begin{array}{l}
x_k(t) = a_k \left[\left(\cos \Omega_k \cos \omega_k - \sin \Omega_k \sin \omega_k \cos i_k \right) \left(\cos \psi_k(t) - e_k \right) \right] \\
+ a_k \left[\left(- \cos \Omega_k \sin \omega_k - \sin \Omega_k \cos \omega_k \cos i_k \right) \sqrt{1 - e_k^2} \sin \psi_k(t) \right] \\
\\
y_k(t) = a_k \left[\left(\cos \Omega_k \cos \omega_k + \cos \Omega_k \sin \omega_k \cos i_k \right) \left(\cos \psi_k(t) - e_k \right) \right] \\
+ a_k \left[\left(- \sin \Omega_k \sin \omega_k + \cos \Omega_k \cos \omega_k \cos i_k \right) \sqrt{1 - e_k^2} \sin \psi_k(t) \right]\\
\\
z_k(t) = a_k \left[\sin \omega_k \sin i_k \left(\cos \psi_k(t) - e_k \right) + \cos \omega_k \sin i_k \sqrt{1 - e_k^2} \sin \psi_k(t) \right] \ , 
\end{array}
\right.
\end{equation}
where $a_k$ is the semi-major axis, $e_k$ the eccentricity, $i_k$ the inclination of the orbit with respect to the ecliptic plane, $\omega_k$ the argument of periapsis, $\Omega_k$ the longitude of ascending node and $\psi_k(t)$ the eccentric anomaly \cite{Boccaletti}.
We chose the initial conditions providing rigid triangular configuration of 
side $l\simeq  5 \times 10^6\ \mbox{km} \simeq 1/30$ AU
\begin{equation}
\left\{
\begin{array}{l}
a_k = 1 \ \mbox{AU}\\
\\
e_k = \left(1 + \frac{2}{\sqrt{3}} \xi + \frac{4}{3} \xi^2 \right)^{1/2} - 1 \\
\\
\tan i_k = \frac{\xi}{1 + \xi/\sqrt{3}} \\
\\
\omega_k = \frac{\pi}{2} \\ 
\\
\Omega_k = \frac{2}{3} \pi (k - 1) - \frac{\pi}{2} \\
\\
M_k(t) = 2 \pi t - \frac{2}{3} \pi (k - 1) - \pi ,
\end{array}
\right.
\label{Initial}
\end{equation}
where $\xi = l/(2 a_k) \simeq 1.67 \times 10^{-2}$ and $M_k(t)$ is the mean anomaly. 
%$l \approx 5 \cdot 10^6$ km is the distance between two spacecrafts and $a = 1$ AU, and $M(t)$ is the mean anomaly. 
Such orbits minimize the variation of the inter-spacecraft distance between the $i$-th and $j$-th 
satellite $\Delta L_{i,j} $ $\equiv |\vec{r}_i(t)-\vec{r}_j(t)|$ \cite{Jiang}, which turns out to be independent of time at the first order in $\xi$ \cite{Dhurandhar}. 
This requirement must be fulfilled in order to operate LISA successfully. 
%We then focused on the relative motions between the couples $(i,j)$ and $(j,l)$, $\delta L_{i,j,l} = \Delta L_{i,j} - \Delta L_{j,l}$, that represent 
In fact, the measured quantities for the GW searches are the differential relative motion between the couples $(i,j)$ and $(j,l)$ of the three reference masses, $\delta L_{i,j,l} = \Delta L_{i,j} - \Delta L_{j,l}$ (see Fig. (\ref{variazione_distanze_imp})). Each couple can be regarded as an unequal arm interferometer and the 
doppler shift induced by the relative differential motions can be measured by a suitable Time Delay Interferometry
(TDI) \cite{TINTO} with a strain sensitivity in Fig. (\ref{sensitivity}) \cite{Bender2}.

%\begin{figure}[!hbt]
%\begin{center}
%\subfloat[]{\includegraphics[width = 7.4 cm, height = 6.1 cm]{dist12.eps}}
%\hspace{3mm}
%\subfloat[]{\includegraphics[width = 7.4 cm, height = 6.1 cm]{diff1223.eps}}
%\caption{a)  Relative motion of 1 and 2 reference masses $\Delta L_{1,2}$ as a function of time. b) Differential 
%relative motion $\delta L_{1,2,3}$ between couples 1-2 and 2-3 of reference masses as a function of time. 
%For different couples, plots are identical but shifted in time, thanks to the principle of equivalence.}
%\label{variazione_distanze_imp}
%\end{center}
%\end{figure}

\begin{figure}[!hbt]
\begin{center}
\includegraphics[width = 8.6 cm, height = 7.2 cm]{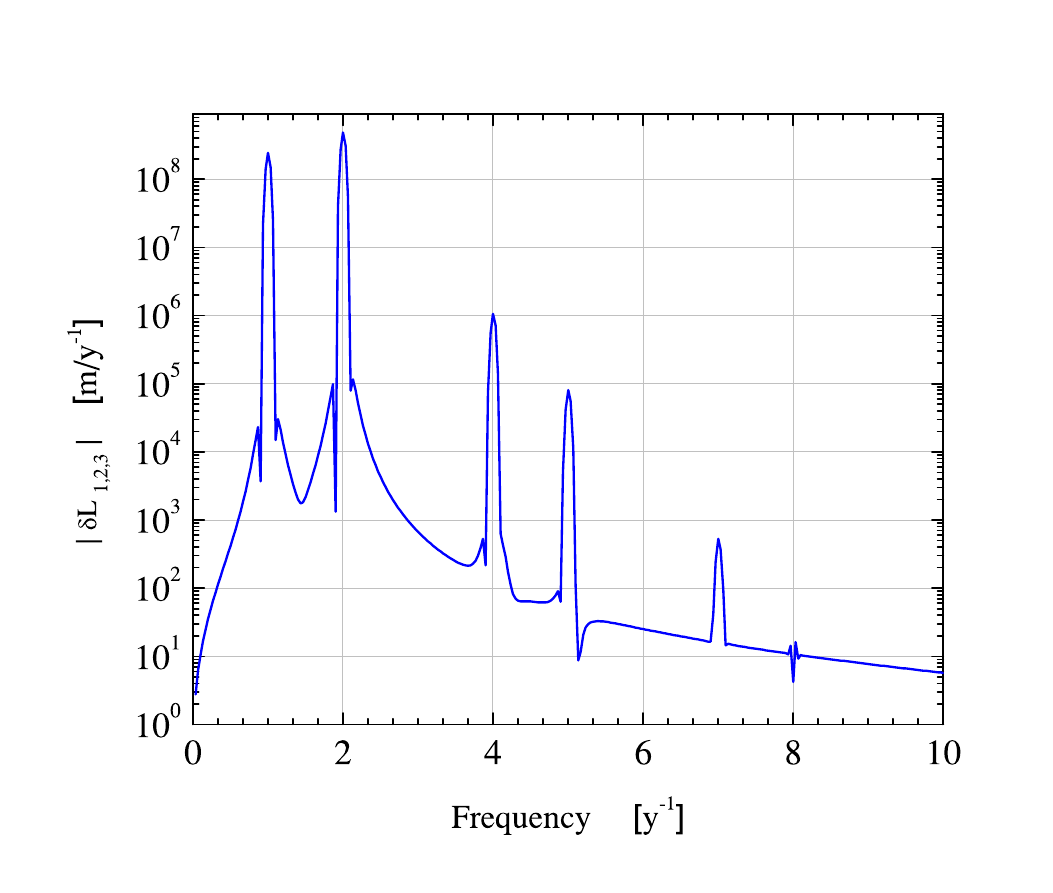}
\caption{$|\widetilde{\delta L}_{1,2,3}(\omega)|$ windowed by a Blackman function with $T=30$ y 
(1 $\mbox{y}^{-1} \approx 3.1709792 \times 10^{-8}$ Hz).  
%Due to the windowing higher harmonics are not visible in the Fourier transform
}
\label{spettriimp}
\end{center}
\end{figure}

The keplerian motions of the LISA reference masses are periodic. We can distinguish their harmonics 
by looking at the modulus of the Fourier transform of $\delta L_{i,j,l}(t)$ \footnote{The Fourier transform of $F(t)$ is as usual $\widetilde{F}(\omega)=
\int_{-\infty}^{+\infty} F(t) \exp(- i \omega t) dt$.}, as in  
Fig. (\ref{spettriimp}), where we have considered a finite observation time $T$ and, to overcome the 
spectral leakage, we have applied a Blackman tapering function \cite{Proakis}. 

%we can determine the harmonics characterizing the periodic motions of the satellites. Since the observation time $T$ will be finite, aspects like spectral leakage must be considered, and the spectrum we present in Fig. (\ref{spettriimp}) was therefore windowed with a \textsl{Blackman function} \cite{Proakis}.

It is worth noticing the absence of the harmonics at 3, 6, 9 \ldots \ $\mbox{y}^{-1}$ due to the 
presence of a discrete symmetry in the equations of motion of the constellation. In fact, 
 as a consequence of the initial conditions in Eq. (\ref{Initial}), the LISA triangle rotates as a ``quasi-rigid body" 
with period of one year and, after integer multiples of one third of year, the dynamical configuration 
is identical to the initial one under any cyclic permutation of the reference mass indices.
However, such a feature is only of theoretical interest, because slightly different initial conditions,
due for instance to the unavoidable injection errors of LISA spacecrafts in their orbits (position $\simeq 2$ Km and velocity $\simeq 2$ mm/s, \cite{Sweetser}), will produce harmonics of the 3 $\mbox{y}^{-1}$ frequency.  

To study the effects of ID on LISA orbits, we apply the perturbation theory in the six-dimensional 
space of parameters and use the Gauss planetary equations. 
Such equations provide time evolution of the orbital parameters under a generic perturbing acceleration field 
$\vec{\gamma}(\vec{r})$ and read
\begin{equation} \label{eqn: gausseq}
\left\{
\begin{array}{l}
\frac{d a_k}{dt} = \frac{2}{n_k \sqrt{1 - e_k^2}} \left[e_k A_r \sin \theta_k + A_t \left(\frac{p_k}{r} \right) \right]\\
\\
\frac{de_k}{dt} = \frac{\sqrt{1 - e_k^2}}{n_k a_k} \left\{ A_r \sin \theta_k + A_t \left[\cos \theta_k + \frac{1}{e_k} \left(1 - \frac{r}{a_k} \right) \right] \right\} \\
\\
\frac{d i_k}{d t} = \frac{1}{n_k a_k \sqrt{1 - e_k^2}} A_n \left(\frac{r}{a_k} \right) \cos (\omega_k + \theta_k)\\
\\
\frac{d \Omega_k}{dt} = \frac{1}{n_k a_k \sin i_k \sqrt{1 - e_k^2}} A_n \left(\frac{r}{a_k}\right) \sin (\omega_k + \theta_k)\\
\\
\frac{d \omega_k}{d t} = - \cos i_k \frac{d \Omega_k}{d t} + \frac{\sqrt{1 - e_k^2}}{n_k a_k e_k} \left[-A_r \cos \theta_k + A_t \left(1 + \frac{r}{p_k} \right) \sin \theta_k \right]\\
\\
\frac{d M_k}{d t} = n_k - \frac{2}{n_k a_k} A_r \left(\frac{r}{a_k} \right) - \sqrt{1 - e_k^2}\frac{d \omega_k}{dt} \ \ ,
\end{array}
\right.
\end{equation}
where $p_k = a_k(1 - e_k^2)$ is the semi-latus rectum, $\theta_k$ is the true anomaly, $n_k = \sqrt{4 \pi^2/a_k^3}$ is the keplerian mean motion and $A_r, A_t$ and $A_n$ are the components of $\vec{\gamma}$ along the versors $\hat{r} = \vec{r}/ \left\| \vec{r} \right\|$ in the radial direction, $\hat{t}$, orthogonal to $\hat{r}$ in the osculating plane and in the direction of $\dot{\vec{r}}$, and $\hat{n} = \hat{r} \times \hat{t}$ \cite{Boccaletti}.

For the generic point $\vec{r} = \left(x, y, z \right)$ close to the ecliptic plane ($i_k \ \approx 10^{-2}$ rad $\forall k$), the accelerating field $\vec{\gamma}(\vec{r})$ has been calculated from the internal gravitational potential of the distribution in Eq. (\ref{eqn: model}), 

$$\Phi(R, z) = \varphi \sum_{n = 0}^{\infty} 
\left(
\begin{array}{c}
1 - \frac{\alpha}{2} \\
n  \\
\end{array}
\right) \times$$ 
\begin{equation}
\times \int_0^{\infty} \left( \frac{R^2}{\tau + a^2}\right)^{1 - \frac{\alpha}{2}} \left[\frac{\tau + a^2}{R^2} \frac{1}{\tau + b^2} \right]^n \frac{z^{2n}}{\left(\tau + a^2 \right) \sqrt{\tau + b^2}} d \tau,
\end{equation}
where $R = \sqrt{x^2 + y^2}$, $\varphi = 2 \pi \rho_0 G r_0^{\alpha} a^{3 - \alpha} \sqrt{1 - \gamma_E^2/\left(1 + \gamma_E^2 \right)} / \left(2 - \alpha\right)$ and $G$ is the Newton gravitational constant. The gravitational potentials of the considered distributions can be easily obtained for the appropriate values of $\alpha$ and $\gamma_E$.

We then numerically integrated the Gauss equations for the three LISA satellites adopting different ID distributions\footnote{We made use of the \emph{Mathematica 7} numerical integrator \emph{NDSolve}; see the documentation webpage http://reference.wolfram.com/mathematica/ref/NDSolve.html}. In order to check the effectiveness of the linear perturbation theory of the studied perturbations, we have compared our results with the numerical solutions obtained by keeping the orbital elements on the RHS unperturbed, and they matched exactly within the integration time (30 years). In addition, the accuracy of the numerical integrations ($\approx 10^{-30}$) have been verified by comparison  with analytical solutions of Gauss equations (approximated to the fourth order in $e$ and with unperturbed orbital parameters on the RHS) for a spherical homogeneous distribution of matter \cite{Cerdonio} and the agreement has been satisfactory. To get the perturbed orbits, and, therefore, the perturbative effects of ID on LISA constellation, we substituted the unperturbed orbital parameters appearing in Eq. (\ref{eqn: orbita}) with the solutions of Eq. (\ref{eqn: gausseq}).

%\begin{figure}[!hbt]
%\begin{center}
%\subfloat{\includegraphics[width = 5.8 cm, height = 4.5 cm]{ell_alpha_a.pdf}}
%\hspace{3mm}
%\subfloat{\includegraphics[width = 5.8 cm, height = 4.5 cm]{ell_alpha_e.pdf}}
%\end{center}
%\end{figure}
%\begin{figure}[!hbt]
%\begin{center}
%\subfloat{\includegraphics[width = 5.8 cm, height = 4.5 cm]{ell_alpha_i.pdf}}
%\hspace{3mm}
%\subfloat{\includegraphics[width = 5.8 cm, height = 4.5 cm]{ell_alpha_nodo.pdf}}
%\end{center}
%\end{figure}
%\begin{figure}[!hbt]
%\begin{center}
%\subfloat{\includegraphics[width = 5.8 cm, height = 4.5 cm]{ell_alpha_omega.pdf}}
%\hspace{3mm}
%\subfloat{\includegraphics[width = 5.8 cm, height = 4.5 cm]{ell_alpha_media.pdf}}
%\caption{Evolution of the orbital parameters for LISA 1; $a_1(0), \ e_1(0), \ i_1(0), \ \omega_1(0), \ \Omega_1(0)$ and $M_1(0)$ are the values of the orbital parameters at $t = 0$, i.e., the unperturbed values.}
%\label{gaussellalpha}
%\end{center}
%\end{figure}

\section{Results}
\label{results}

\begin{figure}[!hbt]
\begin{center}
\includegraphics[width = 8.5 cm, height = 7.3 cm]{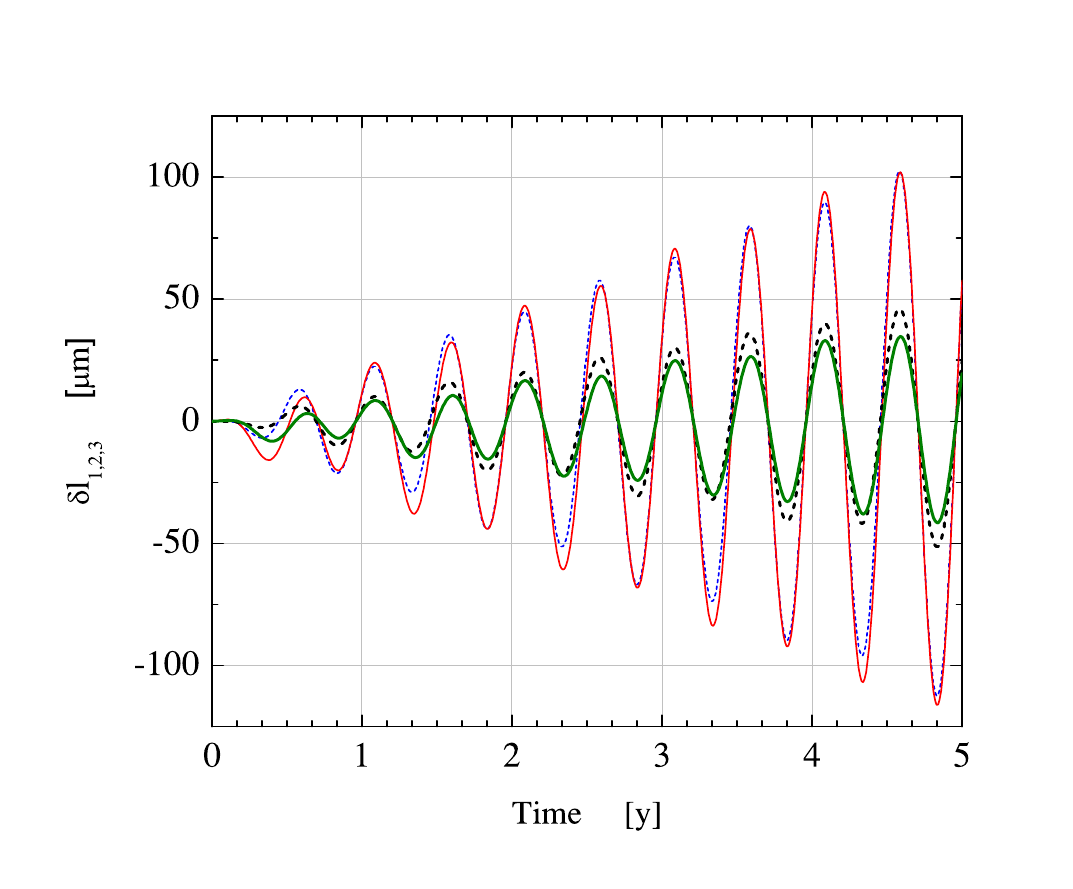}%{dust1223.eps}
\caption{(color on line) Time evolution of $\delta l_{1,2,3}$, for the spherical ($\gamma_E = 0$) distribution (thin dashed line, $\alpha = 0$, thin solid line, $\alpha = 1.3$) and ellipsoidal ($\gamma_E = \sqrt{3}$) (thick dashed line, $\alpha = 0$, thin solid line, $\alpha = 1.3$) distribution. It is worth noticing that ID induces effects of the order of $10^2 \ \mu\mbox{m}$, $10^{12}$ times smaller than those due to the keplerian differential motions.}
\label{diffmotellalpha}
\end{center}
\end{figure}

In Fig. (\ref{diffmotellalpha}), we report the plot representing 
\begin{equation}
\delta l_{1,2,3} = \delta L_{1,2,3}^{pert} - \delta L_{1,2,3}^{unpert},
\end{equation}
i.e., the difference between perturbed and unperturbed differential motions of the constellation for the four considered distributions.

\begin{figure}[!hbt]
\begin{center}
\includegraphics[width = 8.8 cm, height = 7.2 cm]{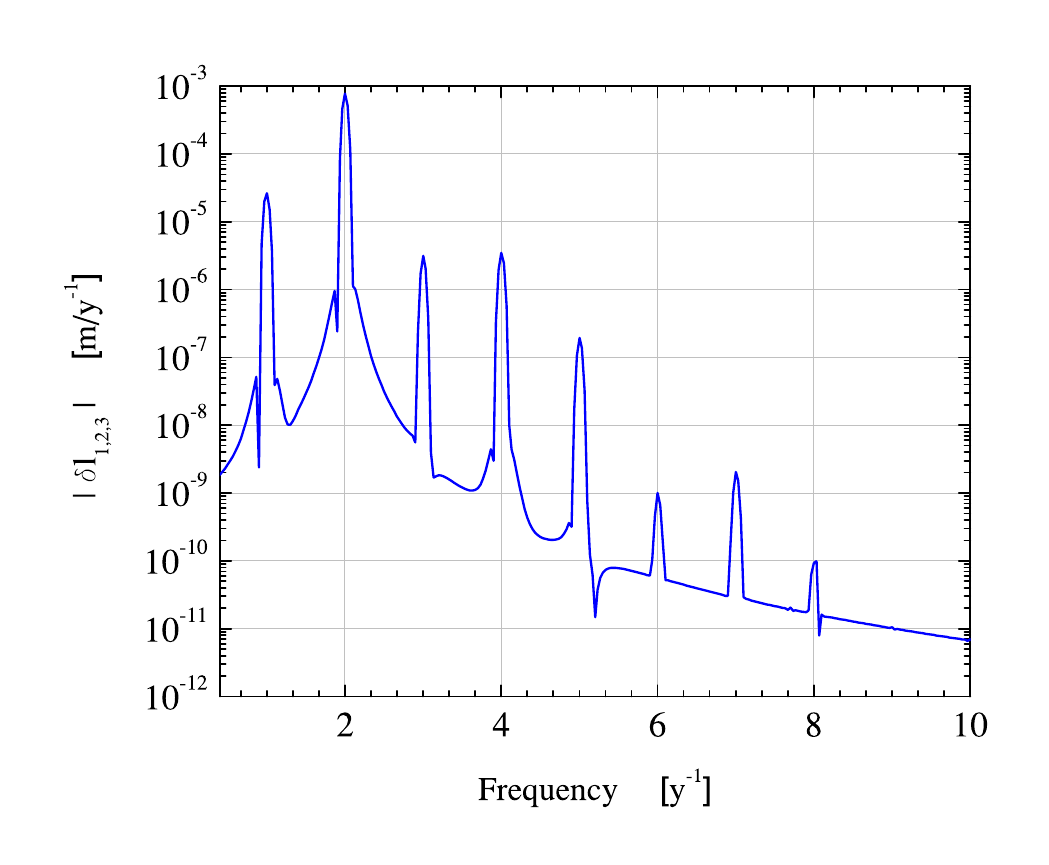}
\caption{Difference between $\delta L_{1,2,3}^{pert}$ and $\delta L_{1,2,3}^{unpert}$ spectra of frequencies after 30 orbits, windowed with Blackman function, obtained with $\alpha = 1.3$ and $\gamma_E = \sqrt{3}$; we note the presence of the harmonics 
corresponding to the integer multiples of 3 $\mbox{y}^{-1}$, contrary to Fig (\ref{spettriimp}).}
\label{diffspectraellalpha}
\end{center}
\end{figure}

We note that the perturbative effects of ID are very small, in fact the LISA constellation opens with an amplitude of the order of $10^2 \ \mu$m after 5 complete orbits. Fig. (\ref{diffspectraellalpha}) shows the modulus of the Fourier transform of the difference between  perturbed and unperturbed differential motions $\delta l_{1,2,3}$, for the ellipsoidal ID distribution with radial profile, and we see that ID enhances resonance peaks, in particular the 2 $\mbox{y}^{-1}$ harmonic. In addition, new harmonics corresponding to integer multiples of 3 $\mbox{y}^{-1}$ appear, as a consequence the ID gravitational field that breaks the permutation symmetry of the equations of motion.  

Therefore, our analysis for the different ID distributions listed at the end of Sec. (\ref{id}) shows that curves in Fig. (\ref{diffmotellalpha}) and Fig. (\ref{diffspectraellalpha}) change their amplitudes of a factor $\approx 2.5$ by varying $\gamma_E$ from $\sqrt{3}$ to 0, while they do not depend significantly on $\alpha$. Additionally, it is easy to show that LISA sensitivity curve is not significantly affected by the ID perturbing effects in the $10^{-4} - 10^{-1}$ Hz frequency band.

\subsection{Optimal ID signal resolution}
\label{resolution}

It is of some interest to investigate the problem of resolving the contributions to the 
differential relative motions due to Sun and ID by means of optimal filtering. 
From the theory of signals resolution we known that if a noisy signal $v(t)$ 
can be a linear combination of two given signals $f(t)$ and $g(t)$ of unknown amplitudes $A$ and $B$
\begin{equation}
v(t) = A f(t) + B g(t) + n(t),
\end{equation}
where $n(t)$ is a zero mean gaussian stochastic process with correlation $\langle n(t)n(t')\rangle = S_0\delta(t-t')$, 
the four following hypothesis are possible:
\begin{enumerate}
	\item $H_0$: neither signal $A$ nor signal $B$ is present;
	\item $H_1$: signal $A$ alone is present;
	\item $H_2$: signal $B$ alone is present;
	\item $H_3$: both signal $A$ and $B$ are present,
\end{enumerate}
which are to be verified by comparing the expectation values of $A$ and $B$, with zero (test of hypothesis). 
The best estimate of $A$ and $B$ turns out to be \cite{Helstrom}
\begin{eqnarray}
\hat{A} &=& \frac{1}{1 - \lambda^2} \int_0^T \left[f(t) - \lambda g(t) \right] v(t) dt\nonumber \\
\hat{B} &=& \frac{1}{1 - \lambda^2} \int_0^T \left[g(t) - \lambda f(t) \right] v(t) dt \ ,
\end{eqnarray}
with standard deviations 

\begin{equation}
\sigma_{\hat{A}} = \sigma_{\hat{B}} = \frac{S_0^{1/2}}{\sqrt{2 \left(1 - \lambda^2 \right)}} \ ,
\end{equation}
where $\lambda = \int_0^T f(t)g(t)dt$, $S_0$ is the noise power density and $f(t)$ and $g(t)$ are normalized to unit energy, $\int_0^T f^2(t) dt = \int_0^T g^2(t) dt = 1.$

\begin{figure}[!hbt]
\begin{center}
\includegraphics[width = 8.3 cm, height = 6.5 cm]{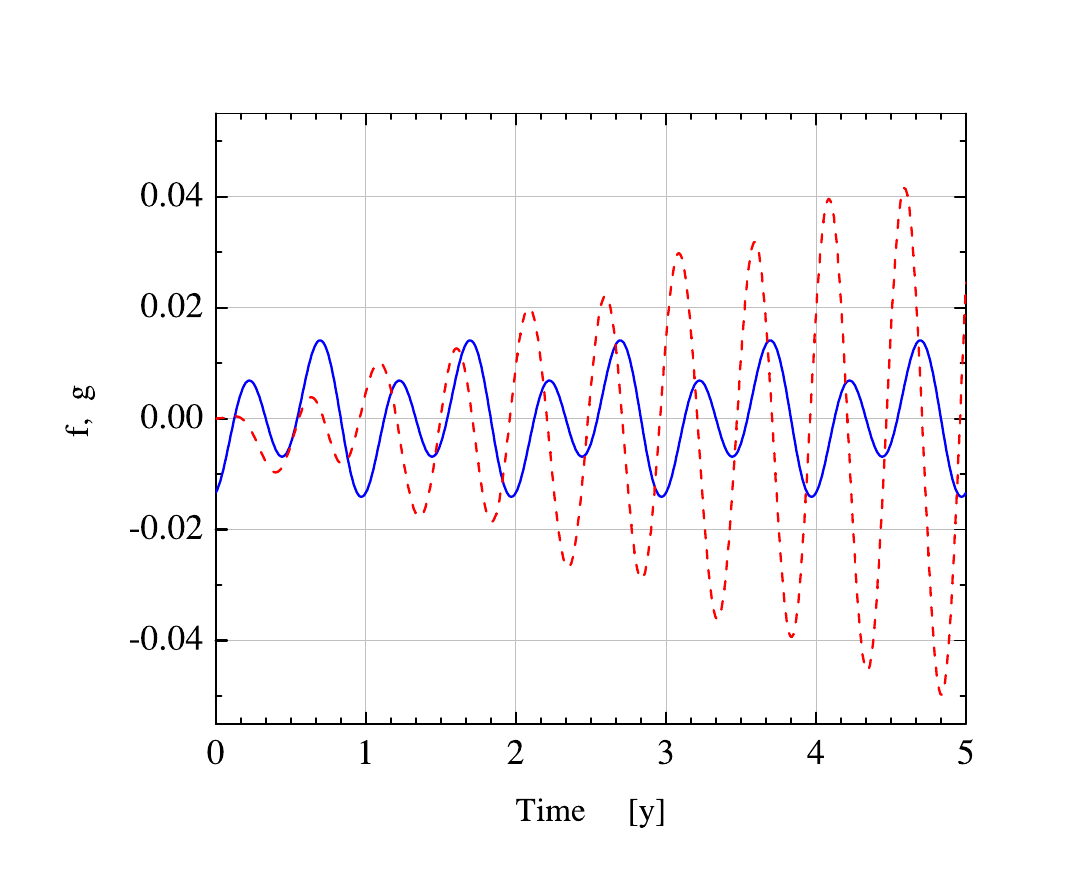}
\caption{(color on line) Time evolution of the two normalized functions $f(t)$ (blue solid line) and $g(t)$ (red dashed line) for the case of an ellipsoidal distribution with radial profile. }
\label{orthogonal}
\end{center}
\end{figure}

In our case, $v(t)$ represents the perturbed differential motions, $Af(t)$ and $Bg(t)$ are the contributes due to Sun and ID respectively. We stress that $f(t)$, $g(t)$, $\hat{A}$ and $\hat{B}$ are almost independent of the initial conditions of LISA reference masses. 
%In addition, we can assume that $H_3$ is true, as the existence of ID is confirmed by observations of zodiacal light. 
It is worth noticing that the two contributions to the perturbed motion are almost orthogonal, 
($\lambda \simeq 10^{-2}$)
as can be seen in Fig. (\ref{orthogonal}), where the normalized functions $f(t)$ and $g(t)$ are plotted.  

As a crude estimation, a strain sensitivity $S_h \approx 10^{-11} \ 1/ \sqrt{\mbox{Hz}}$ at the 2 $\mbox{y}^{-1}$ $ (\approx 6.4 \cdot10^{-8}$ Hz) frequency (the most powerful ID harmonic), is required to achieve a unitary signal-to-noise ratio (SNR = $\hat{A}/\sigma_{\hat{A}} = \hat{B}/\sigma_{\hat{B}}$.) in a 5 years observation time. However, in the literature there exist extrapolations of the strain sensitivity $S_h$ only up to $10^{-6}$ Hz \cite{Bender2} and the measure of such a small signal amplitude depends on the noise level of LISA at very low frequencies ($ 10^{-7}$ Hz), which is still largely unknown.

%We are aware that the detection of ID effects on the differential relative motions of LISA reference masses requires further investigation as noise behaviour is largely unknown in the very low frequency band.

\section{Discussion}
\label{disc}

We have calculated the ID perturbing effects on the LISA differential motions. 
Using the estimated ID density, we found a continuous opening of the LISA orbits of the order of $10^2 \ \mu$m after 5 years at $2\ \mbox{y}^{-1}$ frequency. Moreover, due to the particular LISA orbits, which are nearly circular and very close to the ecliptic plane, we found similar effects for the distributions we have taken into account. In fact, the ID tidal forces vary significantly over a scale much larger than the LISA triangle sides.

All the results we presented for ID also hold in the presence of LDM. In fact, we have shown that the numerical solutions of Eq. (\ref{eqn: gausseq}), for the LISA orbits, are almost independent on $\alpha$ and $\gamma_E$ parameters within the investigated range and of the injection errors. In addition, the perturbations on LISA relative differential motions $\delta L_{i,j,k}$ scale linearly with $\rho_0$.
Thus we can account for LDM just by a rescale of $\rho_0$: for instance, if we assume that the LDM density value is close to the average galactic dark matter (GDM) density value, $\rho_{0,GDM} = 5 \cdot 10^{-22} \ \mbox{kg/m}^3$ as obtained from the galaxy rotation curve, the effect of DM is expected to be 0.5\% that of ID.

%The deviation from the unperturbed motion, $Bg(t)$, will be attributable to a total distribution of mass, $\rho_{tot} \ \propto \ \rho_0$, which includes the two static, non-radiative gravitational fields in the Solar System.
As a consequence, LISA could provide interesting upper limits on both $\rho_{0,ID}$ and $\rho_{0,LDM}$, depending on low frequencies noise and $\rho_{0}$ value, by means of direct gravitational field measurements  instead of observations of zodiacal light in the case of ID. At present time, the best upper limits $\rho_{0,LDM} < 3 \cdot 10^{-16}$ kg/$\mbox{m}^3$ are based on the study of the precession of the perihelions of Mercury, Earth and Mars \cite{Khirplovich}.  
%(, insensitive to the presence of LDM.     

%, which could be better than those obtained through 

However, we stress that a thorough study of LISA displacement noise  below $1 \ \mu$Hz, including local gravitational field fluctuations, thruster noise, orbit determination and injection errors \cite{Bik} etc., is required to establish the relevance of the ID effects in LISA physics.     

%if we call $\Delta$ the ratio of the (unknown) ID and LDM distribution functions:
%\begin{equation}
%\Delta = \frac{\rho_{ID}}{\rho_{LDM}},
%\end{equation} 
%we can therefore write
%\begin{equation}
%\rho_{LDM} = \frac{\rho_{tot}}{1 + \Delta} < \rho_{tot}.
%\end{equation}
%i.e., a measure of the deviations from the keplerian orbits might provide upper limits on $\rho_{LDM}$. As an example, if we assume spherical and homogeneous distributions for both ID and LDM, i.e., those providing the worst upper limits among the considered models, with densities $\rho_{ID} = 10^2 \cdot \rho_{LDM} = 9.6 \cdot 10^{-20} \ \mbox{kg/m}^3$, we then have $\rho_{tot} \approx \rho_{ID}$ and, therefore,
%\begin{equation}
%\rho_{LDM} < 10^{-19} \ \mbox{kg/m}^3,
%\end{equation} 
%which is $\approx 10^{-3}$ times the value obtained through the study of the precession of the perihelions of Mercury, Earth and Mars \cite{Khirplovich}. Different choices of the distribution models provide new values for the upper limits within a factor 20\%. 

\section{Conclusions}

The study of the Solar System gravitational field acting on LISA reference masses is of some relevance also for the detection of GWs. In fact, the non-radiative gravitational perturbations on LISA keplerian motions must be subtracted at the highest accuracy from the relative differential motion in order to measure the GW contributions. In this paper we have calculated the time evolutions of the orbital parameters due to some ID distributions and estimated the opening induced on the LISA constellation,  $\approx 10^2 \ \mu$m after 5 years. The Fourier transform of the perturbation of the differential relative motion has shown an enhancement of the resonances characterizing the unperturbed spectra, in particular the peak at $2\ \mbox{y}^{-1}$ frequency. As a consequence of the LISA orbits, such effects are similar for the 
studied distributions of ID and do not affect the LISA sensitivity band for GW detection. 

On the other hand, the discrimination of very small perturbations from the keplerian differential relative motion depends crucially on the low frequency displacement noise, that in the ideal case should be $\simeq 10^{-11}\ 1/\sqrt{\mbox{Hz}}$ at $2\ \mbox{y}^{-1}$ frequency to detect ID at the density $\sim 10^{-20}\ \mbox{kg/m}^3$ measured by the zodiacal light. 
Unfortunately, a reliable estimate of this noise level at very low frequency is still lacking and requires 
futher investigations. 

Finally, we investigate the possibility of constraining the LDM density with LISA. Even though gravitational 
perturbations  
due to ID and LDM are undistinguishable, the study of the deviations from the keplerian orbits of LISA reference masses could provide interesting upper limits on LDM density.

\section{Acknowledgments}

MS was supported by the Swiss National Science Foundation and the Dr. Tomalla Foundation. GM was supported by the Italian INFN (Istituto Nazionale di Fisica Nucleare). MC and GM are grateful to Professor G. Benettin for useful discussions.


\begin{thebibliography}{9}

\bibitem{TINTO} Armstrong, J.W. {\it et al.}: 2003, `Time delay interferometry' \textbf{\emph{Class. Quantum Grav.}} 20, 283-289.


\bibitem{LISA} Bender, P. {\it et al.}: 1998, `Pre-Phase A Report', second edition.


\bibitem{Bender} Bender, P. {\it et al.}: 2000, `LISA: a cornerstone mission for the observation of gravitational waves System and Technology Study', Report ESA-SCI 11.
 

\bibitem{Bender2} Bender, P.: 2003, `LISA sensitivity below 0.1 mHz', \textbf{\emph{Class. Quantum Grav.}} 20, 301-310.


\bibitem{Bik} Bik J.J.C.M. {\it et al.}: 2007, `LISA satellite formation control', \textbf{\emph{Advances in Space Research}} 40, 25-34.


\bibitem{Boccaletti} Bocaletti, D. and Pucacco, G.: 2001, \textbf{\emph{Theory of orbits}} Springer, Berlin.


\bibitem{Cerdonio} Cerdonio, M. \textsl{et al.}: 2009, `Local Dark Matter searches with LISA', \textbf{\emph{Class. Quantum Grav.}} 26, 094022.


\bibitem{Dhurandhar} Dhurandhar, S.V. \textsl{et al.}: 2005, `Fundamentals of the LISA stable flight formation', \textbf{\emph{Class. Quantum Grav.}} 22, 481-487.


\bibitem{Giese} R.H. Giese \textsl{et al.}: 1986, `Three-dimensional models of the zodiacal dust cloud: a comparative study', \textbf{\emph{Icarus}} 68, 395-411.


\bibitem{Grun} E. Gr\"un  \textsl{et al.}: 1985, `Collisional balance of the meteoritic complex', \textbf{\emph{Icarus}} 62, 244-272.


\bibitem{Helstrom} Helstrom, C.W.: 1960, \textbf{\emph{Statistical theory of signal detection,}} Pergamon Press, New York.

\bibitem{Jiang} Jiang, F. \textsl{et al.}: 2007, `Approximate analysis for relative motion of satellite formation flying in elliptical orbits', \textbf{\emph{Celestial Mech Dyn Astr}} 98, 31-66.


\bibitem{Khirplovich} Khirplovic, I.B. and Pitjeva, E.V.: 2006, `Upper limits on density of dark matter in Solar System', \textbf{\emph{Int. J. Mod. Phys.}} 15, 616-618.


\bibitem{Levasseur} Levasseur-Regourd, A.C., 1996, `Optical and Thermal Properties of Zodiacal Dust", \textbf{\emph{Physics, Chemistry and Dynamics of Interplanetary Dust}}, ASP Conference series 104, 301-308.


\bibitem{Proakis} Proakis, J.G. and Manolakis, D.G.: 1992 \textbf{\emph{Digital signal processing}} Macmillan Publishing Company, second edition, New York.


\bibitem{Sereno} Sereno, M. and Jetzer, Ph.: 2006, `Dark matter versus modifications of the gravitational inverse-square law: results from planetary motion in the Solar system' \textbf{\emph{Mont. Not. R. Astron. Soc.}} 371, 626-632.


\bibitem{Sweetser} Sweetser, T.H.: 2005, `An end-to-end description of the LISA mission', \textbf{\emph{Class. Quantum Grav.}} 22, 429-435. 




























%\bibitem{note1} This is correct only in Newtonian physics: in fact, DM and ID are expected to be gravitationally bound to the Galaxy and to the Solar System respectively; therefore, while gravitoelectric effects are identical for the two, that would not be the case for the gravitomagnetic ones, i.e., effects depending on the speed of the considered objects.

%\bibitem{note2} The distribution is not symmetric with respect to the ecliptic plane, but to the Laplace plane, i.e., the one defined by the total angular momentum of the Solar System. The differences between the two planes are negligible for our scopes.

%\bibitem{newton} In this paper we neglect post-Newtonian and relativistic effects.

%\bibitem{FT} The Fourier transform of $F(t)$ is as usual $\widetilde{F}(\omega)=
%\int_{-\infty}^{+\infty} F(t) \exp(- i \omega t) dt$. 

%\bibitem{note3} We made use of the \textsl{Wolfram Mathematica 7} numerical integrator, http://www.wolfram.com/products/mathematica/index.html.

\end{thebibliography}
\end{document}